\documentclass[12pt]{article}
\usepackage{a4wide}
\usepackage{epsfig}
\usepackage{amsmath}

\newcommand{\half}{{\textstyle\frac{1}{2}}}

\newlength{\absize}
\catcode`@=11
\def\citer{\@ifnextchar [{\@tempswatrue\@citexr}{\@tempswafalse\@citexr[]}}

\def\@citexr[#1]#2{\if@filesw\immediate
  \write\@auxout{\string\citation{#2}}\fi
  \def\@citea{}\@cite{\@for\@citeb:=#2\do
    {\@citea\def\@citea{--\penalty\@m}\@ifundefined
       {b@\@citeb}{{\bf ?}\@warning
       {Citation `\@citeb' on page \thepage \space undefined}}%
\hbox{\csname b@\@citeb\endcsname}}}{#1}}
\catcode`@=12

\begin{document}
  \thispagestyle{empty}
  \pagestyle{empty}
  \renewcommand{\thefootnote}{\fnsymbol{footnote}}
\newpage\normalsize
    \pagestyle{plain}
    \setlength{\baselineskip}{4ex}\par
    \setcounter{footnote}{0}
    \renewcommand{\thefootnote}{\arabic{footnote}}
\newcommand{\preprint}[1]{%
  \begin{flushright}
    \setlength{\baselineskip}{3ex} #1
  \end{flushright}}
\renewcommand{\title}[1]{%
  \begin{center}
    \LARGE #1
  \end{center}\par}
\renewcommand{\author}[1]{%
  \vspace{2ex}
  {\Large
   \begin{center}
     \setlength{\baselineskip}{3ex} #1 \par
   \end{center}}}
\renewcommand{\thanks}[1]{\footnote{#1}}
\begin{flushright}
\end{flushright}
\vskip 0.5cm

\begin{center}
{\large \bf Perturbative Equivalent Theorem in $q$-Deformed Dynamics}
\end{center}
\vspace{1cm}
\begin{center}
Jian-zu Zhang$^{a,b, \S}$
\end{center}
\vspace{1cm}
\begin{center}
$^a$  Department of Physics,
University of Kaiserslautern, PO Box 3049, D-67653  Kaiserslautern, Germany \\
$^b$ Institute for Theoretical Physics, Box 316,
East China University of Science and Technology, 
Shanghai 200237, P. R. China
\end{center}
\vspace{1cm}

\begin{abstract}
Corresponding to two ways of realizing the $q$-deformed Heisenberg algebra 
by the undeformed variables there are two $q$-perturbative Hamiltonians
with the additional momentum-dependent interactions, 
one originates from the perturbative expansion of the  potential, 
the other  originates from that of the kinetic energy term.
At the level of operators, these two $q$-perturbative Hamiltonians are 
different. 
In order to  establish a reliable foundation of  the perturbative
 calculations in $q$-deformed dynamics,
except  examples of the harmonic-oscillator and the Morse potential
demonstrated before,
 the general $q$-perturbative equivalent theorem is  demonstrated, 
which states that for any regular potential which is singularity free
the expectation values of two
 q-perturbative Hamiltonians in the eigenstates of the undeformed 
 Hamiltonian are equivalent.
For the  $q$-deformed ``free'' particle case, the perturbative  Hamiltonian  
originated from the kinetic energy term still keeps its general expression,
 but it does not lead to energy shift.
\end{abstract}

\begin{flushleft}
${^\S}$ E-mail address:  jzzhang@physik.uni-kl.de  \\
\end{flushleft}
\clearpage

The ordinary quantum mechanics, which is based on the Heisenberg 
commutation relation, has obtained every successes from
the space scale  $10^{-8}$ cm to $10^{-18}$ cm. 
There is a possibility that the Heisenberg commutation relation 
at short distances, say, much smaller than  $10^{-18}$ cm, may need 
generalizing. 
In search for such possibility the $q$-deformed Heisenberg algebra is a 
candidate. 
In literature different frameworks of $q$-deformed quantum mechanics were 
established \citer{Schwenk,ZO01}.
The framework of the $q$-deformed Heisenberg algebra developed 
in Refs.~\cite{Hebecker,Fichtmuller} shows 
clear physical content: its relation to the corresponding
$q$-deformed boson commutation relations and the limiting process of
the $q$-deformed harmonic oscillator to the undeformed one are clear. 
In this framework the new features of $q$-deformed
quantum mechanics are explored.
 The $q$-deformed uncertainty relation shows 
essential deviation from the Heisenberg one \cite{JZZ99,OZ00}: 
the ordinary minimal uncertainty relation is undercut.
A non-perturbative feature of the $q$-deformed 
Schr\"odinger equation is that the energy
spectrum exhibits an exponential structure \cite{LW,Fichtmuller,JZZ00}, 
which  qualitatively explains the pattern of quark and lepton masses 
 \cite{JZZ00}. The perturbative expansion of the $q$-deformed Hamiltonian
possesses a complex structure, which amounts to some additional 
momentum-dependent interaction 
\cite{Hebecker,Fichtmuller,JZZ00}.

   Recent studies of the perturbative aspects of the $q$-deformed 
Schr\"odinger equation in the above framework explored interesting 
characteristics.
Corresponding to two ways of realizing the $q$-deformed Heisenberg algebra 
by the undeformed variables which are related by the canonical transformation 
there are two $q$-perturbative Hamiltonians, 
one originates from the perturbative expansion of the  potential, 
the other  originates from that of the  kinetic energy term. 
At the level of operators, these two $q$-perturbative Hamiltonians are 
different. Studies in  the harmonic-oscillator potential 
and the Morse potential showed that \cite{ZO01} expectation values of these 
two q-perturbative Hamiltonians in the eigenstates of the undeformed 
 Hamiltonian are equivalent.
In the example of the harmonic-oscillator potential, two q-perturbative 
Hamiltonians only differ by terms $a^ma^{\dagger n}$ with $m\ne n,$
 thus lead to the same energy shifts 
$\Delta E^{(q)}_n
=-\frac{f^2\omega}{48}\left(4n^3+6n^2+20n+9\right),$
where $a$ and $\omega$ are the annihilation operator and the frequency of 
 the harmonic-oscillator.

In order to establish a reliable foundation of  the perturbative 
calculations in $q$-deformed dynamics  one should clarify: 
whether such equivalence explored in these two examples holds for the 
general case ?
In this letter we demonstrate that for any regular potential  which is 
singularity free the expectation values
 of these two q-perturbative Hamiltonians in the eigenstates of the 
undeformed  Hamiltonian are equal. this is summarized as the perturbative 
equivalent theorem in $q$-deformed dynamics. 
The equivalent theorem means that at the level of operators, 
these two perturbative Hamiltonians are different, however, they differ only 
by a quantity whose expectation value in the undeformed stationary states
 vanishes.
As a self-consistent  check we consider the  $q$-deformed ``free'' particle. 
In this case the  perturbative Hamiltonian  originated from the potential 
vanishes, 
but the other one originated from the kinetic energy term still keeps its 
general expression. 
The calculation confirms that in this case the energy shift is zero. 

   In the following, we first review the necessary background of $q$-deformed 
quantum mechanics.
In terms of $q$-deformed phase space variables --- 
the position operator $X$ and the momentum operator $P$, 
the following $q$-deformed Heisenberg algebra has been
developed \cite{Hebecker, Fichtmuller}:
\begin{equation}
\label{Eq:q-algebra}
q^{1/2}XP-q^{-1/2}PX=iU, \qquad    
UX=q^{-1}XU, \qquad
UP=qPU,
\end{equation}
where $X$ and $P$ are hermitian and $U$ is unitary:
$X^{\dagger}=X$, $P^{\dagger}=P$, $U^{\dagger}=U^{-1}$.
Compared to the Heisenberg algebra the operator $U$ is a new member, 
called the scaling operator. 
The necessity of introducing the operator $U$ is as follows. 

The simultaneous hermitian of  $X$ and  $P$ is a delicate point in 
 $q$-deformed dynamics.
The definition of the algebra 
(\ref{Eq:q-algebra}) is based on the definition of the hermitian momentum 
operator $P$. 
However, if $X$ is assumed to be a hermitian operator in a Hilbert space,
 the usual quantization rule $P\to -i\partial_X$ does not yield a hermitian 
momentum operator. A hermitian momentum operator $P$ is related to 
$\partial_X$ and $X$ in a nonlinear way by introducing a scaling operator 
$U$ \cite{Fichtmuller}
\begin{equation}
\label{Eq:scaling}
U^{-1}\equiv q^{1/2}[1+(q-1)X\partial_X], \qquad
\bar\partial_X\equiv -q^{-1/2}U\partial_X, \qquad
P\equiv -\frac{i}{2}(\partial_X-\bar\partial_X),
\end{equation}
where $\bar\partial_X$ is the conjugate of $\partial_X$. 
The operator $U$ is introduced in the definition of the hermitian momentum, 
thus it closely relates to properties of dynamics and plays an 
essential role in $q$-deformed quantum mechanics. The nontrivial properties 
of  $U$ imply that the algebra (\ref{Eq:q-algebra}) has a richer structure
than the Heisenberg commutation relation. 
In (\ref{Eq:q-algebra}) the parameter $q$ is a fixed real number. 
It is important to make distinctions for different realizations of 
the $q$-algebra by different ranges of $q$ values \citer{Zachos,Solomon}. 
Following Refs.~\cite{Hebecker,Fichtmuller}  we only consider the case $q>1$
 in this paper. The reason is that such choice of  the parameter $q$
leads to consistent dynamics.
In the limit $q\to 1^+$ the scaling operator $U$ reduces 
to the unit operator, thus the algebra (\ref{Eq:q-algebra}) reduces to 
the Heisenberg commutation relation. 

Such defined hermitian momentum $P$ leads to $q$-deformation effects, 
which exhibit in the dynamical equation. 
Eq. (\ref{Eq:scaling}) shows that the  momentum $P$ depends  non-linearly
on $X$ and $\partial_X$. Thus the $q$-deformed 
Schr\"odinger equation is difficult to treat. 
In this letter we demonstrate that there is a reliable foundation for 
 its perturbative calculation.

The $q$-deformed phase space variables $X$, $P$ and 
the scaling operator $U$ can be realized in terms of two pairs of 
 the undeformed variables \cite{Fichtmuller}. 

(I) The variables $\hat x$, $\hat p$ of the ordinary quantum mechanics, 
where $\hat x$, $\hat p$ satisfy:
$[ \hat x, \hat p ]=i$, $\hat x=\hat x^{\dagger}$, 
 $\hat p=\hat p^ {\dagger}$.
The variables $X$, $P$ and 
the scaling operator $U$ are related to $\hat x$, $\hat p$ by:
\begin{equation}
\label{Eq:P-p}
X= \frac{[\hat z+\half]}{\hat z+\half}\hat x,  \qquad
P=\hat p, \qquad
U= q^{\hat z}, \qquad    \hat z=-\frac{i}{2}(\hat x\hat p+\hat p\hat x)
\end{equation}
where $[A]$ is the $q$-deformation of $A$, defined by 
$[A]=(q^A-q^{-A})/(q-q^{-1})$.
 It is easy to check that $X$, $P$ and $U$ satisfy (\ref{Eq:q-algebra}). 

(II) The variables $\tilde x$ and $\tilde p$ of an undeformed algebra, 
which are obtained by a canonical transformation 
of $\hat x$ and $\hat p$:
\begin{equation}
\label{Eq:tilde}
\tilde x=\hat x F^{-1}(\hat z), \qquad \tilde p= F(\hat z)\hat p,
\end{equation}
where 
\begin{equation}
\label{Eq:F(z)}
F^{-1}(\hat z)= \frac{[\hat z-\half]}{\hat z-\half}. \qquad
\end{equation}
Such defined variables $\tilde x$ and $\tilde p$ also satisfy the undeformed 
algebra: $[ \tilde x, \tilde p ]=i$, and 
$\tilde x=\tilde x^{\dagger}$,\quad$\tilde p=\tilde p^{ \dagger}$.
Thus $\tilde p=-i\partial_{\tilde x}$.
The $q$-deformed variables $X$, $P$ and 
the scaling operator $U$ are related to  $\tilde  x$ and $\tilde p$ as follows:
\begin{equation}
\label{Eq:X-x}
X=\tilde x, \qquad P=F^{-1}(\tilde z) \tilde p, \qquad
U= q^{\tilde z}, \qquad  \tilde z=-\frac{i}{2}(\tilde x\tilde p + 
\tilde p\tilde x),
\end{equation}
where $F^{-1}(\tilde z)$ defined by Eq.~(\ref{Eq:F(z)}) for the variables  
($\tilde  x$, $\tilde p$).
From Eqs.~(\ref{Eq:tilde})--(\ref{Eq:X-x}) it follows that such defined $X$,
$P$ and $U$ also satisfy (\ref{Eq:q-algebra}), and Eq. (\ref{Eq:X-x})  is 
equivalent to Eq. (\ref{Eq:P-p}).

The $q$-deformed phase space ($X$, $P$) governed by the $q$-algebra 
(\ref{Eq:q-algebra}) is a $q$-deformation of the phase space 
($\hat x$, $\hat p$) of the ordinary quantum mechanics,
 thus all machinery of  the ordinary quantum mechanics can be applied to
 the $q$-deformed quantum mechanics. 
It means that dynamical equations of the quantum system are the same 
for the undeformed phase space variables ($\hat x$, $\hat p$) and 
for the $q$-deformed  phase space variables ($X$, $P$),
that is,  the $q$-deformed Hamiltonian with potential $V(X)$ is
\begin{equation}
\label{Eq:q-hamiltonian}
H(X,P)=\frac{1}{2\mu}P^{2}+V(X).
\end{equation}

In the  ($\hat x$, $\hat p$) system  $X$ is a non-linear function of
 ($\hat x$, $\hat p$).
From (\ref{Eq:P-p}) it follows that $X$ can be represented as
\begin{equation}
\label{Eq:X-variable}
X=i(q-q^{-1})^{-1} \bigl( q^{(\hat z+1/2)} -
q^{-(\hat z+1/2)}\bigr)\hat p^{-1}. 
\end{equation}
Using (\ref{Eq:X-variable}) it is convenient to discuss the perturbative 
expansion of $X$.
In view of every success of the ordinary quantum mechanics the effects of
$q$-deformation must be extremely small. So we can  let $q=e^{f}=1+f$, 
with $0<f\ll1$. To the order $f^2$ of the perturbative expansion,  $X$ 
reduces to 
\begin{equation}
\label{Eq:X-perturbative}
X=\hat x  + f^2 g(\hat x, \hat p), \qquad
g(\hat x,\hat p)=-\frac{1}{6}(1+
\hat x  \hat p  \hat x  \hat p )\hat x.
\end{equation} 
For any regular potential $V(X)$, which is singularity free, to the 
order $f^2$,  such potentials can be expressed by the 
undeformed variables ($\hat x$, $\hat p$) as
\begin{equation}
\label{Eq:q-potential}
V(X)=V(\hat x) +\hat H^{(q)}_I(\hat x,\hat p),
\end{equation}
with the perturbation
\begin{equation}
\label{Eq:H-q}
\hat H^{(q)}_I(\hat x,\hat p)=f^2\sum_{k=1}^\infty \frac{V^{(k)}(0)}{k!}
\biggl( \sum_{i=0}^{k-1}\hat x^{(k-1)-i}g(\hat x,\hat p)\hat x^{i}\biggr), 
\end{equation}
where $V^{(k)}(0)$ is the $k$-th derivative of $V(\hat x)$ at $\hat x=0.$ 
In (\ref{Eq:H-q}) the ordering between the non-commutative quantities
$ \hat x$ and $g(\hat x,\hat p)$  is  carefully considered. 
Substituting for $g(\hat x,\hat p)$ and summing over $i$ and $k$,
the above result can be expressed as
\begin{equation}
\label{Eq:H-q-summed}
\hat H^{(q)}_I(\hat x,\hat p)
=\frac{f^2}{6}
\bigl\{ \hat x^3 V'(\hat x) \partial^2_{\hat x} 
+ \bigl[\hat x^3 V^{''}(\hat x) +3 \hat x^2 V^{'}(\hat x)\bigr]\partial_{\hat x}
+{\textstyle\frac{1}{3}}\hat x^3 V^{'''}(\hat x) 
+{\textstyle\frac{3}{2}}\hat x^2 V^{''}(\hat x)\bigr\}.
\end{equation}

In the  ($\tilde x$, $\tilde p$) system  $P$ is 
a non-linear function of ($\tilde x$, $\tilde p$).
Using (\ref{Eq:X-x}), to the order $f^2,$ the perturbative expansions of
the momentum $P$ and the kinetic energy $P^{2}/(2\mu)$ read 
\begin{equation}
\label{Eq:P-perturbative}
P=\tilde p  + f^2 h(\tilde x ,\tilde p),\qquad
h(\tilde  x,\tilde p)
=-\frac{1}{6}(1+
\tilde  p\tilde x\tilde  p\tilde x )\tilde p,
\end{equation} 
\begin{equation}
\label{Eq:Ek-perturbative}
\frac{1}{2\mu}P^{2}=\frac{1}{2\mu}\tilde p^{2}+
\tilde H^{(q)}_I(\tilde x,\tilde p), 
\end{equation} 
with
\begin{eqnarray}
\label{Eq:tilde-Hq}
\tilde H^{(q)}_I(\tilde x,\tilde p)&=& \frac{1}{2\mu}f^2
\bigl[\tilde p\, h(\tilde  x,\tilde p)
+h(\tilde  x,\tilde p)\, \tilde p\bigr] 
\nonumber\\
&=& -\frac{1}{12\mu}f^2 \bigl[ 2\tilde x^2 \partial_{\tilde x}^4+
8\tilde x  \partial_{\tilde x}^3+
3\partial_{\tilde x}^2 \bigr]
\end{eqnarray}
Eqs.~(\ref{Eq:Ek-perturbative}) and (\ref{Eq:tilde-Hq}) show that in the 
($\tilde x$,$ $ $\tilde p$) system the perturbative contribution comes 
from the  kinetic-energy term, which is different from Eq.~(\ref{Eq:H-q}),
 where in the ( $\hat x$, $\hat p$) system the perturbative contribution 
comes from the potential.

From Eqs.~(\ref{Eq:q-hamiltonian}),(\ref{Eq:X-perturbative}),
 (\ref{Eq:q-potential}), and (\ref{Eq:H-q-summed}) -
(\ref{Eq:tilde-Hq}),
it follows that the perturbative expansion of the $q$-deformed Hamiltonian 
$H(X,P)$ can be written down in the  $(\hat x,\hat p)$ system 
or in the $(\tilde  x,\tilde p)$ system. In  the  $(\hat x,\hat p)$ system 
\begin{equation}                         
\label{Eq:hat-H}  
H(X(\hat x,\hat p),P(\hat x,\hat p))=H_{\rm un}(\hat x,\hat p)+
\hat H^{(q)}_I (\hat x,\hat p).
\end{equation}
In the  $(\tilde  x,\tilde p)$ system
\begin{equation}                         
\label{Eq:tilde-H}  
H(X(\tilde x,\tilde p),P(\tilde x,\tilde p))=H_{\rm un}(\tilde x,\tilde p)
+\tilde H^{(q)}_I (\tilde x,\tilde p).
\end{equation}
In the above
\begin{equation}                          
\label{Eq:xi-Hun}  
H_{\rm un}(\xi,\kappa)=\frac{1}{2\mu}\kappa^{2}+V(\xi)
\end{equation}
is the corresponding undeformed Hamiltonian in the $(\xi,\kappa)$ system,
where $(\xi,\kappa)$ represents 
$(\hat x,\hat p)$ or $(\tilde  x,\tilde p)$. 

The above two perturbative Hamiltonian $\hat H^{(q)}_I (\hat x,\hat p)$ and 
$\tilde H^{(q)}_I (\tilde x,\tilde p)$ originate,  separately, from the 
perturbative expansions  of the potential and the kinetic energy.
At the level of operator they are different.
 It is interesting to note that their contributions to the perturbative
 shifts of energy spectrum for the undeformed  Hamiltonian in the
 $(\hat x,\hat p)$ system and the $(\tilde  x,\tilde p)$  system  are the same.
 This is summarized in the following theorem.

{\bf Perturbative Equivalent Theorem}:  For any regular potential  which 
is singularity free the expectation
 values of two q-perturbative Hamiltonians $\hat H^{(q)}_I (\hat x,\hat p)$ 
and $\tilde H^{(q)}_I (\tilde x,\tilde p)$ defined,  separately, by 
 Eqs.~(\ref{Eq:H-q-summed}) and (\ref{Eq:tilde-Hq}), 
 in the eigenstates of the undeformed  Hamiltonian are equivalent.

Suppose that the Schr\"odinger equation for the undeformed system is solved in 
the configuration space $\xi_0$, i.e., the manifold of the spectrum $\xi_0$ of 
$\xi$:
\begin{equation}                          
\label{Eq:Hun-xi}  
H_{\rm un}(\xi,\kappa)\psi_n^{(0)}(\xi_0)=
E^{\rm (un)}_n\psi_n^{(0)}(\xi_0).
\end{equation}
The structure of $\psi_n^{(0)}(\hat x_0)$ in the configuration space 
$\hat x_0$ and the structure of $\psi_n^{(0)}(\tilde x_0)$ in the 
configuration space $\tilde x_0$ are the same.

The  expectation values  of $\hat H^{(q)}_I$ and $\tilde H^{(q)}_I$ 
  in the undeformed stationary states $\psi^{(0)}_n$ are
\begin{eqnarray}                          
\label{Eq:En}  
\Delta \hat E^{(q)}_n &=& \int d\hat x_0 \psi_n^{(0)\ast}(\hat x_0 )
\hat H^{(q)}_I(\hat x_0,-i\partial_{\hat x_0} )\psi_n^{(0)}(\hat x_0 ) 
\nonumber\\
\Delta\tilde E^{(q)}_n &=& \int d\tilde x_0 \psi_n^{(0)\ast}(\tilde x_0)
\tilde H^{(q)}_I(\tilde x_0,-i\partial_{\tilde x_0}) \psi_n^{(0)}(\tilde x_0),
\end{eqnarray}  
 From Eqs.~(\ref{Eq:H-q-summed}) and (\ref{Eq:tilde-Hq}), 
using the Schr\"odinger equation and integrating Eq.~(\ref{Eq:En})  by parts, 
it follows that
 Eq.~(\ref{Eq:En}) can be rewritten as
\begin{equation}
\label{Eq:Delta-E-hat}
\Delta \hat E^{(q)}_n
=\frac{f^2}{6}\int_{-\infty}^\infty d\hat x_0 \psi^{(0)*}_n(\hat x_0)
\Bigl\{V(\hat x_0) \Bigl[1-4\mu\hat x_0 ^2 \Bigl(V(\hat x_0) 
-E^{\rm (un)}_n\Bigr)\Bigr]
-{\textstyle\frac{2}{3}}\mu E^{\rm(un)}_n  \hat x_0^3 V'(\hat x_0)\Bigr\}
\psi^{(0)}_n(\hat x_0).
\end{equation}
\begin{equation}
\label{Eq:Delta-E-tilde}
\Delta\tilde E^{(q)}_n
=\frac{f^2}{6}
\int_{-\infty}^\infty d\tilde x_0\, \psi^{(0)*}_n(\tilde x_0)
\Bigl(V(\tilde x_0)- E^{\rm (un)}_n\Bigr) \Bigl[ 1-
4\mu \tilde x_0^2\Bigl(V(\tilde x_0)-
E^{\rm (un)}_n\Bigr) \Bigr]\psi^{(0)}_n(\tilde x_0).
\end{equation}
Using the Schr\"odinger equation again,  because 
the structure of $\psi_n^{(0)}(\hat x_0)$ in the configuration space 
$\hat x_0$ and the structure of $\psi_n^{(0)}(\tilde x_0)$ in the 
configuration space $\tilde x_0$ are the same,
 the difference of
$\Delta \hat E^{(q)}_n$ and $\Delta\tilde E^{(q)}_n$  is given by
\begin{equation}
\label{Eq:Delta-Delta-E}
\Delta \hat E^{(q)}_n - \Delta\tilde E^{(q)}_n
=  \frac{f^2}{6}E^{\rm (un)}_n 
\int_{-\infty}^\infty dx \psi^{(0)*}_n(x)
\Bigl[1-2 x^2 \partial_x^2
-{\textstyle\frac{2}{3}} x^3\mu V'(x)\Bigr]\psi^{(0)}_n(x).
\end{equation}
In  the undeformed stationary states $|\psi^{(0)}>$ we have
\begin{equation}                          
\label{Eq:Virial}
i\frac{d}{dt} \langle \psi^{(0)}|\hat x^m\hat p^n|\psi^{(0)}\rangle
= \langle \psi^{(0)}|\left[ \hat x^m\hat p^n, 
\frac{1}{2\mu} \hat p^2+V(\hat x )\right]|\psi^{(0)}\rangle=0.
\end{equation}
From Eq.~(\ref{Eq:Virial}) for the cases of $m=3,$ $n=1$ and 
$m=2,$  $n=0$ it follows that 
\begin{equation} 
\langle \psi^{(0)}|\hat x^3 V'-\frac{3}{\mu}\hat x^2\hat p^2
+\frac{3i}{\mu}\hat x \hat p |\psi^{(0)}\rangle=0, \qquad
\langle \psi^{(0)}|1+2i\hat x \hat p|\psi^{(0)}\rangle = 0. \nonumber 
\end{equation}
Putting these two equations in Eq.~(\ref{Eq:Delta-Delta-E}), it shows 
$\Delta \hat E^{(q)}_n - \Delta\tilde E^{(q)}_n =0.$

As a consistent check of the  equivalent theorem we consider the  
$q$-deformed ``free'' particle described by the Hamiltonian 
$H(X,P)=\frac{1}{2\mu}P^2$.
In this case $\hat H^{(q)}_I(\hat x,\hat p)=0$, but 
$\tilde H^{(q)}_I(\tilde x,\tilde p)$ is still described by 
Eq.~(\ref{Eq:tilde-Hq}). The question is whether in this case
 $\Delta\tilde E^{(q)}_n=0$ ? In the eigenstate $|\psi_p^{(0)}\rangle$
of the undeformed free Hamiltonian
 $H_{\rm un}(\tilde x,\tilde p)=\frac{1}{2\mu}\tilde p^{2}$, from
Eq.~(\ref{Eq:Virial}) for the cases of $m=n=3$ and $m=n=2$  it follows that
\begin{equation}
\langle \psi_p^{(0)}|i\tilde x^2\tilde p^4
+ \tilde x\tilde p^3 |\psi_p^{(0)}\rangle =0, \qquad
\langle \psi_p^{(0)}|2i\tilde x\tilde p^3
+ \tilde p^2 |\psi_p^{(0)}\rangle =0. \nonumber
\end{equation}
Putting these results in Eq.~(\ref{Eq:tilde-Hq}), we obtain 
 $\Delta\tilde E^{(q)}_n = \langle \psi_p^{(0)}|
 \tilde H^{(q)}_I(\tilde x,\tilde p)|\psi_p^{(0)}\rangle = 0.$

It should be pointed out that if $q$-deformed quantum mechanics is
 a realistic physical theory, its effects mainly manifest at very short
 distances much smaller than $10^{-18}$~cm; 
its correction to the ordinary quantum mechanics must be extremely small
 in the energy range of nowadays experiments, 
which means that the parameter $q$ must be very close to one.
So the perturbative investigation of  $q$-deformed dynamics is meaningful,
 which shows the  clear indication of $q$-deformed modifications 
to the ordinary quantum mechanics, and in some interesting cases, for example 
in the $q$-squeezed state \cite{OZ00},  may provide some evidence about 
such effects to nowadays experiments.

The equivalent theorem establishes  a reliable foundation for  the
 perturbative calculations in  $q$-deformed dynamics.
  Base on the equivalent theorem we can use any one of  two $q$-perturbative 
Hamiltonians to calculate the energy  shifts. For systems  with 
complicated potentials, it is convenient to calculate 
the $q$-perturbative shifts of the energy spectrum in  the 
$(\tilde x,\tilde p)$ system.

\vspace{0.4cm}
  This work has been supported by the Deutsche Forschungsgemeinschaft 
(Germany). The project started during the author's visit at 
Max-Planck-Institut f\"ur Physik (Werner-Heisenberg-Institut).
 He would like to thank Prof. J. Wess, Prof. H.J.W. M\"uller-Kirsten
 and Prof. P. Osland for stimulating discussions.
 His work has also been supported by the National Natural Science 
Foundation of China under the grant number 10074014 and by the Shanghai 
Education Development Foundation. 


\end{document}